\documentclass[aps,twocolumn,prl,superscriptaddress,preprintnumbers,showpacs,tightenlines]{revtex4}
\usepackage{amssymb}
\usepackage{epsfig,graphicx,times}

\begin{document}

\preprint{First Draft}
\title{Universal Existence of Exact Quantum State Transmissions in
Interacting Media }
\author{Lian-Ao Wu}
\affiliation{Ikerbasque--Basque Foundation for Science and Department of Theoretical
Physics and History of Science, The Basque Country University (EHU/UPV), PO
Box 644, 48080 Bilbao, Spain}
\affiliation{Advanced Science Institute, The Institute of Physical and Chemical Research
(RIKEN), Wako-shi, Saitama 351-0198, Japan}
\author{Yu-xi Liu}
\affiliation{Advanced Science Institute, The Institute of Physical and Chemical Research
(RIKEN), Wako-shi, Saitama 351-0198, Japan}
\affiliation{CREST, Japan Science and Technology Agency (JST), Kawaguchi, Saitama
332-0012, Japan}
\author{Franco Nori}
\affiliation{Advanced Science Institute, The Institute of Physical and Chemical Research
(RIKEN), Wako-shi, Saitama 351-0198, Japan}
\affiliation{CREST, Japan Science and Technology Agency (JST), Kawaguchi, Saitama
332-0012, Japan}
\affiliation{Center for Theoretical Physics, Physics Department, Center for the Study of
Complex Systems, The University of Michigan, Ann Arbor, Michigan 48109-1120}
\date{\today}

\begin{abstract}
We consider an exact state transmission, where a density matrix in one
information processor A at time $t=0$ is exactly equal to that in another
processor B at a later time. We demonstrate that there always exists a
complete set of orthogonal states, which can be employed to perform the
exact state transmission. Our result is very general in the sense that it
holds for arbitrary media between the two processors and for any time
interval. We illustrate our results in terms of models of spin, fermionic
and bosonic chains. This complete set can be used as bases to study the
perfect state transfer, which is associated with degenerated subspaces of
this set of states. Interestingly, this formalism leads to a proposal of
perfect state transfer via adiabatic passage, which does not depend on the
specific form of the driving Hamiltonian.
\end{abstract}

\maketitle

\bigskip \pagenumbering{arabic}

\emph{Introduction.--- }One of the central missions in quantum information
theory is to transmit known or unknown quantum states from one region to
another, for instance, from one information processor A to another B. The
transmission may be a two quantum-state swap, short-distance communications
between components of a quantum device, or long-distance quantum
communications through optical fibers. Flying photons have been considered
as the main candidate for information carrier or channel in quantum
communications. However, the simplest form of transmission, two-state swap
between spin qubits, may be processed in terms of local spin couplings, such
as the direct Heisenberg interaction. In this case, interactions play the
role of information carriers. Indeed, quantum state transmissions through
interaction-linked chains \cite{Bose,Christandl}, such as spin chains and
bosonic lattices, have been investigated extensively (see, e.g., \cite%
{Bose07} and references therein). Early works concentrated on the
transmitting abilities of the naturally-available interactions of spin
chains, but in most cases it failed to perfectly transmit a quantum state 
\cite{Bose}. Afterwards \cite{Christandl,Bose07,Burgarth,Nikolopoulos}
interactions were proposed where perfect quantum state transfer was possible.

In spite of many works in this area, a central question remains open: given
an evolution operator $U(\tau )$, governed by a time-independent or
dependent Hamiltonian $H(t)$ linking the two processors A and B, is it
possible that there exist states, during an arbitrary time interval $\tau $,
being transmitted exactly from region (or processor) A at time $t=0$ to
region (processor) B at $t=\tau $ ? Here we first define an exact state
transmission in the sense that a \textit{pure or mixed} density matrix in
region (or processor) A and at time $t=0$ is transmitted exactly to another
region (processor) B (with the same internal structure as A) at the time $%
\tau $. We can thus show that there always exists a complete set $\left\{
\Psi _{k}(0)\right\} $ of orthogonal states, which can be used to perform
the exact state transmission. The $\Psi _{k}(0)$'s are states of the entire
system, which refers to two processors and also the media between them.
Throughout this paper, the phrase \textit{exact state transmission} refers
specifically to the fact that the density matrices at two processors are
equal because of the use of the set $\left\{ \Psi _{k}(0)\right\} $. Our
result is very general in the sense that it holds for arbitrary media
between the two processors and for an arbitrary time $\tau $ (where, of
course, $\tau $\ is inside the light cone). We illustrate the set $\left\{
\Psi _{k}(0)\right\} $ for models of spin, as well as spinless fermionic and
bosonic chains. Indeed, the perfect state transfer (PST) in \cite{Christandl}
occurs in a degenerate subspace of $\left\{ \Psi _{k}(0)\right\} .$ Based on
this result, we propose an approach for perfect state transfer via adiabatic
passage, which does not rely on specific form of the driving Hamiltonian.

\emph{Universal existence of exact state transmission.---} We consider an $M$
dimensional system ($M$ can be infinite), e.g., $M=2^{K}$ for $K$ qubits,
located in processor A, spanned by the bases $\left\{ \left\vert \alpha
\right\rangle \right\}.$ A \emph{pure or mixed} quantum state in this processor
can be generically characterized by a density matrix $\rho ^{A}(0)$. We want
to transmit this state to another processor B at a given time interval $\tau 
$. The exact state transmission is defined by 
\begin{equation}
\rho ^{A}(0)=\rho ^{B}(\tau )  \label{exact}
\end{equation}%
or by their matrix elements $\rho _{\alpha \beta }^{A}(0)=\rho _{\alpha
\beta }^{B}(\tau ).$ Here $\rho ^{B}(\tau )$ describes the quantum state of
processor B at the given time $\tau .$ Assume that the entire system,
composed of processors A, B, and the media, is initially in the state $\Psi
(0).$ We show below that there exists a complete orthogonal set $\left\{
\Psi _{k}(0)\right\} _{\tau },$ which depends on $\tau $, such that \emph{an
exact state transmission described by Eq.}~(\ref{exact})\emph{\ occurs if
the initial state is one of the states in this set}.

Proof: For the initial state $\Psi (0),$ the matrix elements of $\rho
^{A}(0) $ in processor A at time $t=0$ are 
\begin{equation}
\rho _{\alpha \beta }^{A}(0)=\left\langle \Psi (0)|\beta _{A}\right\rangle
\left\langle \alpha _{A}|\Psi (0)\right\rangle ,  \label{a}
\end{equation}%
where $\alpha ,\beta =1,2,...,M$ and $\left\vert \beta _{A(B)}\right\rangle $
refers to a state $\left\vert \beta \right\rangle $ in processor A (B). At a
given time $t=\tau ,$ the matrix elements of $\rho ^{B}(\tau )$ in processor
B are%
\begin{eqnarray}
\rho _{\alpha \beta }^{B}(\tau ) &=&\left\langle \Psi (0)\right\vert
U^{\dagger }(\tau )\left\vert \beta _{B}\right\rangle \left\langle \alpha
_{B}\right\vert U(\tau )\left\vert \Psi (0)\right\rangle  \label{b} \\
&=&\left\langle \Psi (0)\right\vert U^{\dagger }(\tau )\mathcal{P}\left\vert
\beta _{A}\right\rangle \left\langle \alpha _{A}\right\vert \mathcal{P}%
U(\tau )\left\vert \Psi (0)\right\rangle  \nonumber \\
&=&\left\langle \Psi ^{\prime }(\tau )|\beta _{A}\right\rangle \left\langle
\alpha _{A}|\Psi ^{\prime }(\tau )\right\rangle ,  \nonumber
\end{eqnarray}%
where $U(\tau )$ is the evolution operator of the entire system (two
processors and the media). Here we also introduce the $A\Leftrightarrow B$
exchange operator $\mathcal{P}$, satisfying $\mathcal{P}\left\vert \beta
_{A}\right\rangle \left\langle \alpha _{A}\right\vert \mathcal{P}=\left\vert
\beta _{B}\right\rangle \left\langle \alpha _{B}\right\vert $ and $\mathcal{P%
}^{2}=1$. The exchange operator $\mathcal{P}$ swaps all the states of two
processors. It can be expressed explicitly by 
\begin{equation}
\mathcal{P}=\sum (\left\vert \beta _{A}\right\rangle \left\langle \alpha
_{A}\right\vert )\otimes (\left\vert \alpha _{B}\right\rangle \left\langle
\beta _{B}\right\vert )  \label{exchange}
\end{equation}%
and changes its form with different bases. The state $\Psi ^{\prime }(\tau
)=G(\tau )\Psi (0)$, where we introduce the operator $G(\tau )=\mathcal{P}%
U(\tau )$. This operator behaves similar (but not equal) to an evolution
operator and here will be called a quasi-evolution operator. It is
significant to note that the operator $G(\tau )$ is unitary and satisfies 
\begin{equation}
G^{\dagger }(\tau )G(\tau )=U^{\dagger }(\tau )\mathcal{PP}U(\tau )=1.
\label{U}
\end{equation}%
As any unitary operator, the operator $G(\tau )$ can be diagonalized and has
a complete set of orthonormal eigenvectors $\left\{ \Psi _{k}(0)\right\}
_{\tau }$ and exponential eigenvalues $\left\{ \exp (i\phi _{k})\right\}
_{\tau }$. A vector $\Psi _{k}(0)$ in the set obeys the eigenequation: 
\begin{equation}
G(\tau )\Psi _{k}(0)=\exp (i\phi _{k})\Psi _{k}(0).  \label{eigen}
\end{equation}%
Comparing the expressions of $\rho _{\alpha \beta }^{A}(0)$ in Eq.~(\ref{a})
with $\rho _{\alpha \beta }^{B}(\tau )$ in Eq.~(\ref{b}), it is easy to
conclude that if the initial state $\Psi (0)$\ is one of the $\Psi _{k}(0)$%
's, the equality $\rho _{\alpha \beta }^{A}(0)=\rho _{\alpha \beta
}^{B}(\tau )$ or Eq.~(\ref{exact}) holds. In other words, an exact state
transmission occurs. $\blacksquare $

The above statement or proposition could be very useful in investigating
state transmissions. Generically, for an arbitrary Hamiltonian and at an
arbitrary time $\tau $, we can numerically diagonalize $G(\tau )$ to obtain
its eigenstates and eigenvalues, especially for small systems. In fact,
studies on short-distance transmission in small systems is more significant
for information exchange between the components of a quantum computing
device. Below we will illustrate several eigenproblems of the operator $%
G(\tau )$ using analytical models. However, we emphasize that \emph{the
proposition is universal and only results from the fact that a unitary
operator possesses a complete set of orthogonal eigenvectors}.

\emph{Fully controllable models.--- } We first consider a one-dimensional
lattice (a chain) with $N$ local units (particles), each has the same (could
be infinite) dimensional eigenspace spanned by the basis $\left\{ \left\vert
s\right\rangle \right\} $, e.g., $s=0,1$ for a spin chain. Assume that we
are able to control the interactions between nearest-neighbor sites $(\,j-1)$
and $j$. We then turn on/off the permutation operators $E_{j,j-1}$ in
chronological order such that $U_{\!f}(\tau )=E_{N,N-1},...E_{32}E_{21},$
where $E_{jj-1}=\sum (\left\vert s_{j}\right\rangle \left\langle
r_{j}\right\vert )\otimes (\left\vert r_{j-1}\right\rangle \left\langle
s_{j-1}\right\vert )$ which, e.g., in a spin chain, can be represented by
the \textit{XY} or Heisenberg interactions \cite{Wu02}. An arbitrary state $%
\left\vert \phi \right\rangle _{1}\left\vert R\right\rangle $ at the first
site as processor A, is an eigenstate of $G_{\!f}(\tau ),$ where $\left\vert
R\right\rangle $ is an arbitrary state for the rest of the chain $%
(j=2,...,N) $.

\emph{Systems with exchange symmetry.---} This is a simple but general case.
If $[U(\tau ),\mathcal{P}]=0$, e.g., in a time-independent system
Hamiltonian $U(\tau )=e^{-iH\tau }$ leading to $[H,\mathcal{P}]=0,$ the
eigenproblem of $G(\tau )$ becomes that of $U(\tau )$ and $\mathcal{P}$.
Specially, the common eigenstates of the time-independent Hamiltonian $H$
and $\mathcal{P}$ may be those of $G(\tau )$ as well.

\emph{Adiabatic Eigenstates.--- }The adiabatic approximation is usually
applied to describe systems under slowly-varying time-dependent Hamiltonians 
\cite{WKB}. The adiabatic quantum state transfer for spin-1 chains was
studied in \cite{Eckert07}. We now consider a time-dependent Hubbard-type
Hamiltonian \cite{Jaksch98}, 
\begin{equation}
H(t)=-J(\Delta )T_{h}+\omega (\Delta )h_{s}+H_{U},  \label{BHH0}
\end{equation}%
where $T_{h}=\sum (a_{j}^{\dagger }a_{j+1}+a_{j+1}^{\dagger }a_{j})$ is the
hopping term. $n_{j}=a_{j}^{\dagger }a_{j}$ is the number operator at site $j
$. The creation operator $a_{j}^{\dagger }$ satisfies the standard
commutation relations for bosons $a_{j}=b_{j}$ and anticommutation relations
for spinless fermions $a_{j}=c_{j}$. This model can also be equivalent to a
general \textit{XY} model in spin chains through the Jordan-Wigner
transformation \cite{Wu02}. The parameter $\Delta =\frac{\tau }{2}-t$; and $%
H_{U}=U\sum n_{j}(n_{j}-1)$ is the on-site repulsion ($H_{U}\equiv 0$ for
fermions). The single particle energy $h_{s}=\sum \epsilon _{j}n_{j}$ is
designed such that $\epsilon _{j}<$ $\epsilon _{j+1}$ for all $j$'$s$ and $%
\omega (\frac{\tau }{2})>0.$ We also require that $J(\Delta )$ [$\omega
(\Delta )$] is an even [odd] function of $\Delta $ and $J(\pm \frac{\tau }{2}%
)=0$ when $t=0(+)$ and $\tau (-)$. Also $H(0)=\omega (\frac{\tau }{2}%
)h_{s}+H_{U}$ and $H(\tau )=-\omega (\frac{\tau }{2})h_{s}+H_{U}.$ In this
case processor A (B) is at the first (last) site. An initial state $%
a_{1}^{\dagger }\left\vert \mathbf{0}\right\rangle $ is the lowest
eigenstate of $H(0)$ when the total particle number is one, where $%
\left\vert \mathbf{0}\right\rangle $ is the bosonic or fermionic vacuum
state. If we control the evolution of $H(t)$ adiabatically from time $0$ to $%
\tau ,$ the final state will be $e^{-i\phi (\tau )}a_{N}^{\dagger
}\left\vert \mathbf{0}\right\rangle $, where $\phi (\tau )$ is the sum of
the dynamic phase angle and geometric phase angle. The state $a_{1}^{\dagger
}\left\vert \mathbf{0}\right\rangle $ satisfies 
\begin{equation}
G_{\text{ad}}(\tau )a_{1}^{\dagger }\left\vert \mathbf{0}\right\rangle
=e^{-i\phi (\tau )}a_{1}^{\dagger }\left\vert \mathbf{0}\right\rangle ,
\label{ad}
\end{equation}%
which is an eigenstate of the quasi-evolution operator $G_{\text{ad}}(\tau )$%
. Another trivial eigenstate of $G_{\text{ad}}(\tau )$ is the vacuum state $%
\left\vert \mathbf{0}\right\rangle .$ For spin or spinless fermionic chains
and total particle number larger than one, the $n$-particle product states
of $c_{1}^{\dagger }c_{2}^{\dagger }...c_{n}^{\dagger }\left\vert \mathbf{0}%
\right\rangle $ are also eigenstates of $G_{\text{ad}}(\tau )$. However, if
the total boson number is larger than one, there are two situations due to
the relative strength of $\omega (\frac{\tau }{2})$ and $U.$ In the limit
when $H_{U}=0,$ the eigenstates of $G_{\text{ad}}(\tau )$ will be condensed
to $c_{1}^{\dagger n}\left\vert \mathbf{0}\right\rangle .$ When the on-site
repulsion is strong, the eigenstates have the same form as those of femionic
chains if the total boson number is not large.

The hopping term $T_{h}$ drives the whole system from site $1$ to site $N$.
It is important to note that the hopping term does not affect the above
formalism, as long as it drives the system within the adiabatic regime. In
other words, this adiabatic protocol is insensitive to the driving
Hamiltonian, which makes it a promising candidate in state transmission. The
approach is applicable to higher dimensional systems.

\emph{Linear model.--- }We now consider a linear Hamiltonian with the same
notations for operators as in (\ref{BHH0}) 
\begin{equation}
H=\sum J_{j}(e^{i\vartheta }a_{j}^{\dagger }a_{j+1}+e^{-i\vartheta
}a_{j+1}^{\dagger }a_{j})+\sum \epsilon _{j}n_{j},  \label{BHH}
\end{equation}%
where $\vartheta $ is a phase angle. It may be a linearly-coupled bosonic
Hamiltonian which is equivalent to that of the on-chip coupled cavities
(e.g., in Refs.~\cite{Hartmann,Bliokh,Zhou08}) or a generalized Bose-Hubbard
model without on-site repulsion \cite{Jaksch98}. The desired parameters $%
\epsilon _{j}$ and $J_{j}$ can be realized by experimental methods, such as
tunable transmission line resonators, SQUID couplers~\cite{Sandberg},
external magnetic traps \cite{Jaksch98}. The Hamiltonian may also represent
a general \textit{XY} model of a spin chain.

A recent work \cite{wupst} discussed the mapping between a rank $l$
irreducible spherical tensor bosonic operator $A_{lm}^{\dagger }$ and the
creation operator $a_{k}^{\dagger }$ at site $k=m+\frac{N+1}{2}.$ Those
results can be used directly to the spinless fermonic and spin cases modeled
by Eq.~(\ref{BHH}). The three components of the angular momentum vector $%
\mathbf{L}$ may be expressed by creation and annihilation operators of
fermions or bosons,%
\begin{eqnarray*}
L_{x} &=&\sum D_{j}(a_{j}^{\dagger }a_{j+1}+a_{j+1}^{\dagger }a_{j}) \\
L_{y} &=&i\sum D_{j}(a_{j}^{\dagger }a_{j+1}-a_{j+1}^{\dagger }a_{j}) \\
L_{z} &=&\sum \left( j-\frac{N+1}{2}\right) n_{j}
\end{eqnarray*}%
where $D_{j}=\frac{\sqrt{j(N-j)}}{2}$. If we select $J_{j}$ and $\epsilon
_{j}$ in the Hamiltonian (\ref{BHH}) such that $J_{j}=JD_{j}$ \cite%
{Christandl} and $\epsilon _{j}=0$ (the case $\epsilon _{j}=\epsilon (j-%
\frac{N+1}{2})$ will be considered in the next section), the time-evolution
operator of the Hamiltonians becomes%
\begin{equation}
U(\tau )=\exp (-iJ\tau L_{x})  \label{U(t)}
\end{equation}%
The evolution operator $U(\tau )$ corresponds to a rotation operator $%
R(\Omega ).$ The irreducible tensor operator $A_{lm}^{\dagger }$ in the
Heisenberg representation evolves as $A_{lm}^{\dagger }(\tau )=\sum (-)^{i%
\frac{\pi }{2}(m^{\prime }-m)}d_{m^{\prime }m}^{l}(-J\tau )A_{lm^{\prime
}}^{\dagger }$. When $\tau =\pi /J,$ the expression is reduced to a simple
form $U^{\dagger }(t_{0})a_{i}^{\dagger }U(t_{0})=ra_{N-i+1}^{\dagger }$,
where the factor $r=\exp (i\pi \frac{N-1}{2})$, and its interesting effect
has been discussed in \cite{wupst}. It is easy to show that, in this case,
we can use $\mathcal{P}=r^{\ast }e^{i\pi L_{x}}$ as an exchange operator,
which is equivalent to the exchange operator within the two processors and
mirror exchanging the site indices. The quasi-evolution operator $G(\tau )$
now becomes%
\begin{equation}
G_{l}(\tau )=r^{\ast }\exp (-i(J\tau -\pi )L_{x})  \label{g1}
\end{equation}%
An eigenstate $\Psi _{k}(0)$ may be expressed by the product of the
operators $\widetilde{A}_{lm}^{\dagger }=\exp (-i\frac{\pi }{2}%
L_{y})A_{lm}^{\dagger }\exp (i\frac{\pi }{2}L_{y})$ acting on the vacuum
state $\left\vert \mathbf{0}\right\rangle $. At the time $\tau =\pi /J,$ the
operator $G_{l}(\pi /J)=r^{\ast }$ is a constant. All product states of $%
a_{i}^{\dagger }$ acting on the vacuum state are the eigenstates of $%
G_{l}(\pi /J)$.

\emph{Generalization using the dressing transformation.---} The
time-independent dressing transformations $W$ preserve the commutation
relations \cite{wupst} among the angular momentum components but introduce
new effects. The whole family of Hamiltonians generated \cite{Wu03} by
dressing transformations $W$ can behave as the linear cases studied in the
last section. We obtain the quasi-evolution operators and their eigenstates
of two models via individual dressing transformations. As an example, under
the transformation $W_{j}=\exp \left[ -\vartheta (j-\frac{N+1}{2})n_{j}%
\right] $, the Hamiltonian $H_{l}=JL_{x}$ \cite{Christandl} becomes $%
H_{l}^{\prime }=J(\cos \vartheta L_{x}+\sin \vartheta L_{y})$. The
eigenstates of $H_{l}^{\prime }$ and the set $\left\{ \Psi _{k}(0)\right\} $
can be expressed by a product of the tensor operators $A_{lm}^{\dagger
}(\vartheta )=\exp (i\vartheta L_{z})A_{lm}^{\dagger }\exp (-i\vartheta
L_{z})$ acting on the vacuum state. The quasi-evolution operator is $%
G_{r}(\tau )=r^{\ast }\exp [-i(J\tau -\pi )\left( \cos \vartheta L_{x}+\sin
\vartheta L_{y}\right) ]$. At the time $\tau =\pi /J,$ the quasi-evolution
operator $G_{r}(\pi /J)=r^{\ast }$ is a constant. Again, all product states
of $A_{lm}^{\dagger }(\vartheta )$ acting on the vacuum state are the
eigenstates of $G_{r}(\pi /J)$. Another example is the one-mode squeezing
transformation $W_{j}=\exp [\xi (b_{j}^{2}-b_{j}^{\dagger 2})/2],$ where the
transformed Hamiltonian reads as $H_{l}^{\prime }=J\left( \cosh \xi
L_{x}+\sinh \xi L_{x}^{\prime }\right) ,$ where $L_{x}^{\prime }=\sum
D_{j}(b_{j}^{\dagger }b_{j+1}^{\dagger }+b_{j+1}b_{j}).$ In this case, the
quasi-evolution operator $G_{s}(\tau )=r^{\ast }\exp [-i(J\tau -\pi )\left(
\cosh \xi L_{x}+\sinh \xi L_{x}^{\prime }\right) ].$ At the time $\tau =\pi
/J,$ again $G_{s}(\pi /J)=r^{\ast }$ is a constant.

\emph{Perfect state transfer.---} Perfect state transfer was found \cite%
{Christandl,Bose07,Burgarth,Nikolopoulos} by designing specific strengths of
the coupling constants in spin chains. Interesting studies of generic
properties of perfect state transfer have been carried out in the last few
years (see, e.g., \cite{Yung,Shi,Nori}). Below we will look into PST in
terms of the complete orthogonal set $\left\{ \Psi _{k}(0)\right\} $.

Although exact state transmissions exist universally, not all of them are
significant for PST. For instance, the vacuum state $\left\vert \mathbf{0}%
\right\rangle $ is an eigenstate of $G(\tau )$ in the above example, such
that $\rho ^{A}(0)=\rho ^{B}(\tau ),$ but there is no actual information
transmitted in the process since the two processors share the same
information on this state. A significant exact state transmission for the
PST requires that at least some of the eigenstates of $G(\tau )$ in the set $%
\left\{ \Psi _{k}(0)\right\} $ are biased to occupy processors A and B.
Ideally, if an eigenstate of $G(\tau )$ is localized at processor A then
that state can be perfectly transferred. Here, localization means that the
targeted density matrix $\rho ^{A}(0)$ is a state that does not entangle
other states outside processor A.

In quantum information theory, quantum state transfer often refers to
transferring an unknown state. Since the set $\left\{ \Psi _{k}(0)\right\} $
is a complete orthogonal set, a known or unknown initial state can be
expanded as\ $\Psi (0)=\sum_{k=1}^{M}C_{k}\Psi _{k}(0)$, which is usually
not an eigenstate of $G(\tau )$ because 
\begin{equation}
G(\tau )\Psi (0)=\sum_{k=1}^{M}C_{k}e^{i\phi _{k}}\Psi _{k}(0).
\label{expand1}
\end{equation}%
At a specific time $\tau ^{\ast }$, $\Psi (\tau ^{\ast })$ may again become
an eigenstate of $G(\tau ^{\ast })$ such that $\Psi (\tau ^{\ast })=e^{i\phi
(\tau ^{\ast })}\Psi (0).$ The condition to satisfy this equation is $%
C_{k}[\exp (i\phi _{k})-\exp (i\phi (\tau ^{\ast })]=0$. Therefore, for $%
C_{k}\neq 0,$ one obtains the condition $\phi (\tau ^{\ast })=\phi _{k}+2\pi
K_{k}$, where $K_{k}$ are arbitrary integers. This is a very restrictive
condition if there are many coefficients $C_{k}\neq 0$, which may only
happen for particular systems with symmetry. As examples, in the above
linear system [when $\tau ^{\ast }=\pi /J,$ because $G_{l}(\pi /J),G_{s}(\pi
/J)$ and $G_{r}(\pi /J)$ are constants,] any state in processor A is an
eigenstate of $G(\tau ^{\ast }).$ However, if there are few nonzero $C_{k}$%
's, the perfect state transfer still happens even without symmetry. The
adiabatic process is an example. Assume that we are initially in a
superposition state $\Psi (0)=C_{1}\left\vert \mathbf{0}\right\rangle
+C_{2}a_{1}^{\dagger }\left\vert \mathbf{0}\right\rangle $, then 
\[
G(\tau )\Psi (0)=C_{1}\left\vert \mathbf{0}\right\rangle +C_{2}e^{-i\phi
(\tau )}a_{1}^{\dagger }\left\vert \mathbf{0}\right\rangle 
\]%
is not an eigenstate of $G(\tau )$ but becomes an eigenstate at a given time 
$\tau ^{\ast }$ when $\phi (\tau ^{\ast })=2\pi .$ Thus, $\Psi (\tau ^{\ast
})$ is an eigenstate of $G(\tau ^{\ast })$. The state evolves from time 0 to 
$\tau ^{\ast }$ such that 
\[
C_{1}\left\vert \mathbf{0}\right\rangle +C_{2}a_{1}^{\dagger }\left\vert 
\mathbf{0}\right\rangle \rightarrow C_{1}\left\vert \mathbf{0}\right\rangle
+C_{2}a_{N}^{\dagger }\left\vert \mathbf{0}\right\rangle . 
\]%
This is a perfect adiabatic state transfer. When the on-site repulsion is
small, an arbitrary function $f(b_{1}^{\dagger })$ can also be perfectly
transferred .

\emph{Perfect mixed-state transfer.--- }If the eigenstate $\Psi (\tau ^{\ast
})$ is not completely localized at processor A, a reduced density matrix at
processor A with matrix elements $\rho _{\alpha \beta }^{A}=\left\langle
\Psi (\tau ^{\ast })|\beta _{A}\right\rangle \left\langle \alpha _{A}|\Psi
(\tau ^{\ast })\right\rangle $, can be perfectly transferred from processor
A to B. As an example, we will consider a \emph{collective} dressing
transformation. Eq.~(\ref{U(t)}) with $\epsilon \neq 0$ can be regarded as
an evolution operator via a dressing transformation $W=\exp (i\theta L_{y}),$
where $\theta =\arctan (\epsilon /J)$, for which we have $H=JL_{x}+\epsilon
L_{z}=\sqrt{J^{2}+\epsilon ^{2}}\,WL_{x}W^{\dagger }$. At the time $\tau
^{\ast }=\pi /\sqrt{J^{2}+\epsilon ^{2}},$ a dressed tensor is $%
A_{lm}^{\dagger }(\theta )=WA_{lm}^{\dagger }W^{\dagger }=\sum d_{mm^{\prime
}}^{l}(\theta )A_{lm^{\prime }}^{\dagger }.$ Products of tensors $%
A_{lm}^{\dagger }(\vartheta )$ acting on the vacuum state, for example $%
A_{lm}^{\dagger }(\vartheta )\left\vert \mathbf{0}\right\rangle $, are
eigenstates of $G(\tau ^{\ast }).$ A state $\Psi (0)=C_{1}\left\vert \mathbf{%
0}\right\rangle +C_{2}A_{l-l}^{\dagger }(\theta )\left\vert \mathbf{0}%
\right\rangle $ cannot be prepared at processor A located at the first site.
The reduced density matrix at site $1$ now becomes 
\begin{equation}
\rho ^{A}=\left[ 
\begin{array}{cc}
\left\vert C_{1}\right\vert ^{2} & C_{1}C_{2}^{\ast }\left( \cos \frac{%
\theta }{2}\right) ^{N-1} \\ 
C_{2}C_{1}^{\ast }\left( \cos \frac{\theta }{2}\right) ^{N-1} & \left\vert
C_{2}\right\vert ^{2}\left( \cos \frac{\theta }{2}\right) ^{2N-2}%
\end{array}%
\right] ,  \label{rhoa}
\end{equation}%
where we have used $d_{-l-l}^{l}(\theta )=\left( \cos \frac{\theta }{2}%
\right) ^{N-1}.$ However, this mixed state can still be perfectly
transferred from site $1$ to site $N$ during the time interval $\tau ^{\ast
} $. The density matrix (\ref{rhoa}) becomes a pure state when $N$ goes to
infinity or $\theta $ is small (i.e., when $\epsilon $ is small, as in
experiments~\cite{Greiner02}). In that case, it becomes a perfect pure-state
transfer. We emphasize here that although we transfer a mixed state, our
transfer is still perfect. Our result is different from previous ones where
the fidelity is less than 100\% (see, e.g., ref. \cite{Mark} and references
therein).

\emph{Conclusion.---} We have shown the general existence of a set of
initial states such that exact state transmissions can take place. The
result is universal in the sense that it holds for arbitrary interactions,
through \textit{any} media between the two processors and at \textit{any}
given evolution time. The existence of such a set is essentially based on
the properties of a unitary operator. We have shown a unitary operator,
called the quasi-evolution operator, whose complete orthogonal set of
eigenstates can perform the exact state transmission. We illustrate the
"transferring power" of this set of eigenstate through analytical models.
Generally the quasi-evolution operator can be numerically diagonalized,
especially for small systems. Quantum information processing requires to
transfer unknown states. The set can be used as a basis to perform perfect
state transfer in its degenerate subspaces. Significantly, the present
formalism leads us to propose an adiabatic perfect state transfer protocol,
which is insensitive to the Hamiltonian driving the transfer.\ \ 

\noindent \emph{Acknowledgments.} This work was supported in part by the
NSA, LPS, ARO, NSF Grant No. EIA-0130383, and JSPS-RFBR No. 06- 02-91200.
LAW has been supported by the Ikerbasque Foundation.

\end{document}